\documentstyle[11pt,paspconf]{article}

\begin{document}

\title{The Magellanic Clouds, Past, Present and Future - A Summary of IAU 
Symposium No.190}
\author{Sidney van den Bergh}
\affil{Dominion Astrophysical Observatory\\
Herzberg Institute of Astrophysics\\
National Research Council of Canada\\
5071 West Saanich Road\\
Victoria, British Columbia\\
Canada V8X 4M6}


\section{Introduction}

Important problems to which we would like to find answers are:
\begin{itemize}
\item{What are the distances to the Large Magellanic Cloud (LMC) and 
the Small Magellanic Cloud (SMC)?}
\item{What is the present distribution of stars, gas and dark matter in the 
Clouds, and how did it evolve?}
\item{How, and where, did the Magellanic Clouds form, and how have their 
orbits evolved?}
\item{Finally the recent discovery of numerous microlensing events in the 
Clouds provides answers to questions that we have only recently started
to ask.}
\end{itemize}

\noindent For background material on the Clouds of Magellan the reader is 
referred to IAU Symposium No.~108 (van den Bergh \& de Boer 1984), IAU 
Symposium No.~148 (Haynes \& Milne 1991), and to the recent monograph 
The Magellanic Clouds (Westerlund 1997).  The sections of this summary 
of the conference proceedings are given approximately in the order in 
which they were presented at the symposium.

\section{Interstellar Matter}

\subsection{Optical Imaging}

A number of authors presented beautiful narrow-band images of both the 
LMC and the SMC, which showed an intricate network of hot bubbles 
produced by the stellar winds of giant associations, and shells formed 
by exploding supernovae.  From the close association between supernova 
remnants (SNRs) and emission nebulosity Petre (NASA/GSFC) concluded 
that the majority (35 out of 48) of the SNRs observed in the Large 
Cloud had been produced by supernovae of Type II (SNe II).  However, 
the fact that the SNR N 103B [which was produced by a SN Ia (Hughes et 
al.~1995)] is associated with an H II region shows that individual type 
assignments based on environment may be incorrect.  In the SMC the 
presently available data are insufficient to derive the relative 
numbers of SNe Ia and SNe II.  It would be of interest to compare 
recent narrow-band images of the Magellanic Clouds with digitized 
versions of similar images obtained a few decades ago, to search for 
the light echoes of prehistoric supernovae in the Clouds.  Sonneborn 
(NASA/GFSFC) et al.~reported on the recent development of a hot spot 
associated with the forward shock produced by SN 1987A, that is now 
beginning to enter the dense gas on the inner-most edge of the ring 
surrounding this supernova.  This may be the beginning of the ``second 
coming'' of SN 1987A!

\subsection{Infrared Radiation}

Beautiful new $^{12}$CO ($\mathrm{J} = 1-0$) observations obtained with 
the NANTEN telescope on Las Campanas were shown by various Japanese groups.  
Due to low metallicity molecular clouds are found to be rarer in the LMC 
than they are in the Galaxy.  Not unexpectedly such clouds are seen to 
be even scarcer in the SMC.  Over the mass range 10$^5$ M$_{\odot}$  to 
10$^6$ M$_{\odot}$ 
Mizuno (Nagoya) et al.~find that these molecular clouds have a power 
law mass spectrum of the form $dN/dM \propto M^{-1.5}$.  Saito (Nagoya) et 
al.~note that the positions of H II regions are strongly correlated with 
those of giant molecular clouds (GMCs).  The positions of young star 
clusters are slightly less correlated with those of GMCs.  Finally 
there is little evidence for a correlation between the positions of 
SNRs and GMCs.  This suggests that most GMCs have lifetimes that are 
short compared to the $\geq 1 \times 10^7$ yr time-scale for the 
evolution of SNe II.

\subsection{21-cm Observations}

A number of interesting new results, based on observations obtained 
with the Australia Telescope Compact Array (ATCA), were presented at 
the meeting.  The neutral hydrogen observations by Dickey (U. 
Minnesota) et al.~show that the cool atomic phase is quite abundant in 
the Clouds.  Possibly such cool gas would have been transformed into 
molecular clouds in the dust-rich Galactic environment.  Alternatively 
the Galactic gas may have been warmed by photoelectric heating of small 
grains.  The hydrogen gas in the LMC forms a well-defined rotating 
disk, whereas that in the SMC appears chaotic.  The LMC rotation curve 
implies a total disk mass of $2.5 \times 10^9$ M$_{\odot}$ within 4 kpc of 
the kinematic 
center of the Large Cloud.  The total amount of neutral gas (including 
He) in this disk is $5 \times 10^8$ M$_{\odot}$.  
Observations with the ATCA show a 
number of features in the outer regions of the LMC that have been 
stretched into transient spiral arm-like structures by differential 
rotation.  From their $\sim 20^{\circ}$ pitch angle Staveley-Smith (ATNF) 
estimates an age of  $\sim 90$ Myr for these features.  The ATCA survey 
shows an intricate system of bubble-like structures in both the LMC and 
SMC.  A few of these bubbles appear to have no central 
cluster/association or supernova remnant.  The origin of such bubbles 
is a mystery.

\subsection{Absorption-line Spectroscopy}

Hot ($\mathrm{T}>10^5$ K) gas is best observed in the ultraviolet absorption 
lines of highly-ionized atoms, such as O$^{+5}$, N$^{+4}$, C$^{+3}$ 
and Si$^{+3}$.  
Such lines may form near O3-O7 stars in the LMC disk, or in a hot gaseous 
corona around the Large Cloud.  Wakker (U. Wisconsin) has been able to 
use the {\em Hubble Space Telescope} (HST) to show that some lines of sight, 
which do not pass close to any early O-type stars, show strong C IV 
absorption.  A comparison of the velocities of such hot gas with that 
of H$\alpha$ emission shows that it is not co-spatial with disk material.  
This suggests that this hot gas is located in a corona around the LMC.  
The LMC is found to contain an extensive distribution of hot plasma 
which radiates  $\sim 10^{38}$ ergs s$^{-1}$ in the 0.5-2.0 keV band.  
The SMC emits only  $\sim 10^{37}$ ergs s$^{-1}$ in this energy band.  
De Boer et al.~(1998) have 
detected H$_2$ in absorption along various lines of sight towards the LMC. 
They derive excitation temperatures
$\leq 50$ K for levels J $\leq 1$ and $\sim 470$ K for levels 
$2 \leq $ J $\leq 4$.  
From these observations they conclude that moderate UV pumping affects 
even the lowest levels of excitation towards the LMC.  Richter et al.~(1998)
have detected an H$_2$ cloud with an excitation temperature  
$\simeq 70$ K 
at  +120 km s$^{-1}$ towards the SMC, while another feature at  
+160 km s$^{-1}$ has an excitation temperature
$> 2300$ K, which must be due to strong UV radiation from its energetic 
environment.

\section{The History of Star Formation in the Clouds}

\subsection{Massive Stars}

The determination of the slope of the mass spectrum of star formation 
in the Clouds of Magellan has a long and sordid history.  Massey (KPNO) 
pointed out that many of the problems encountered by previous 
investigators had been due to the fact that they only used photometric 
data do determine the slope of the mass function $\Gamma$.  Massey finds that 
much more consistent results can be obtained by combining spectroscopic 
and photometric data on massive stars.  He finds that 
$\Gamma = -1.4 \pm 0.2$
in the LMC, $\Gamma = -1.3 \pm 0.1$ in the SMC, and 
$\Gamma = -1.3 \pm 0.4$ in the 
Milky Way.  These results suggest that the mass spectrum of star 
formation may be universal.  Massey called attention to the fact that 
$\Gamma = -1.4 \pm 0.1$ in the super-compact cluster R 136, which is 
situated at the center of 30 Doradus complex.  
This value is identical to that for 
other regions of star formation in which the stellar density is two 
orders of magnitude lower.  Walborn (STScI) pointed out that 
populations of differing age can be recognized in the 30 Dor region:  
(1) The central ionizing cluster R 136, which has an age of 2-3 Myr, 
(2) a younger generation of early-O stars and IR sources that are 
embedded in bright nebular filaments to the west and northeast of R 
136, (3) an older population of late-O and early-B supergiants, having 
an age of 4-6 Myr that is scattered throughout the 30 Dor region, (4) a 
still older compact cluster 3' northwest of R 136 that contains A and M 
supergiants having ages of $\sim 10$ Myr, and (5) a 4-6 Myr old  
association, that includes the luminous blue variable R 143 in the 
southern part of the nebula.  In other words it appears that star 
formation in the 30 Dor region is an ongoing process, rather than a 
single event.

Langer \& Heger (U. Potsdam) have compared theoretical predictions of 
the frequency distribution of massive evolved stars (Hess diagram) with 
observations and conclude that (1) theory yields a gap to the right of 
the main sequence that is not observed, and (2) theory predicts a 
dependence of [Fe/H] on the ratio of luminous blue stars to M-type 
supergiants that is in the opposite sense to that which is actually 
observed.  They suggested that these problems might be resolved by 
using both [Fe/H] {\em and} rotation in stellar models.  Such rotational 
mixing will destroy the onion-like internal structure of non-rotating 
stellar models.  In the discussion following their presentation Massey 
suggested that both (1) and (2) might be partly accounted for by 
observational selection effects.

Element ratios are potentially important sources of information on 
evolutionary history.  Smith (U. Texas) showed that the abundance ratio 
of the r-process element europium to that of the s-process element 
barium is higher in the Magellanic Clouds, than it is for stars of 
similar metallicities in the Galaxy.  Since the mass spectrum of star 
formation appears to be universal this effect cannot be due to 
differences in the frequency with which SNe II having different 
progenitor masses occur.

A problem of long standing (Richtler, Spite \& Spite 1991, and 
references therein) is that stars in the young SMC cluster NGC 330 
appeared to be more metal deficient than similar stars in the SMC 
field.  Such a difference would be difficult to explain.  However, new 
work by Hill  (Paris Obs.) appears to show that the abundance 
differences between NGC 330 and the SMC field, and between NGC 1818 and 
NGC 2100, and the LMC field are not significant.

De Boer et al.~(1997) have recently proposed that star formation may 
be triggered as gas is compressed in a bow-shock formed at the leading 
edge of the LMC.  Such a scenario would favor star formation at the SE 
edge of the Large Cloud, where the cumulative effects of the space 
motion and rotation of the Large Cloud are greatest.  However, a recent 
study of the distribution of 2138 supergiants and 1170 Cepheids by 
Grebel (Lick Obs.) \& Brandner (JPL) does not appear to support the 
bow-shock scenario.  This suggests that the Galactic halo gas density 
is negligible at R$_{\mathrm Gc} \sim 50$ kpc.

\subsection{History of Star Formation in the Large Cloud}

Observations of field stars in the LMC by Butcher (1977) showed that a 
burst of star formation, that has continued to the present day, started 
in the Large Cloud 3-5 Gyr ago.  This conclusion has been supported by 
all subsequent studies.  However, the magnitude of this increase in the 
rate of star formation, that occurred $\sim 4$ Gyr ago, remains 
controversial.  Using HST observations in three outer fields Geha et 
al.~(1998) find that the rate of star formation increased by only about 
a factor of three.  Gallagher (U. Wisconsin) detected no evidence for 
significant differences between the evolutionary histories in different 
LMC fields.  However, Romaniello (STScI/Pisa) et al.~find that the rate 
of star formation in the field surrounding SN  1987A is presently an 
order of magnitude higher than it was $\sim 5$ Gyr ago.  On balance, it 
appears that available data from Large Cloud field stars (Geha et al.~1998) 
indicate that half of all LMC stars formed during the last 4 Gyr, 
with the other half having formed during the preceding $\sim 10$ Gyr.  Da 
Costa (MSSSO) showed that this contrasts dramatically with the 
situation for star clusters in the Large Cloud.  Elaborating on earlier 
work (Da Costa 1991) showed that the rate of cluster formation in the 
LMC was close to zero between $\sim 4$ Gyr ago, and the era of globular 
cluster formation $\sim 14$ Gyr ago.  Taken at face value his data suggest 
that the rate of cluster formation had increased by almost two orders 
of magnitude $\sim 4$ Gyr ago.  This conclusion was strengthened by 
Sarajedini  (1998) who found only 3 new clusters with ages $\sim 5$ Gyr in 
the $\sim 10$ Gyr age gap between the era of globular cluster formation and 
the burst that started $\sim 4$ Gyr ago.  The existence of these clusters 
might indicate that the burst ``ramped-up'' for $\sim 1$ Gyr before reaching 
maximum intensity $\sim 4$ Gyr ago.  
{\em The dramatic contrast between the 
history of cluster formation and that of field stars suggests that star 
clusters cannot be used as proxies for star formation.}  
In fact, it 
would have been very difficult to understand how stars with elevated 
[Fe/H] values could have formed $\sim 4$ Gyr ago if star (and supernova!) 
formation rates had remained depressed during the preceding ``dark 
ages''. The observation that the Local Group galaxy IC 1613 is forming 
young stars but few (if any) star clusters (Baade 1963, van den Bergh 
1979), provides a clear demonstration of the fact that the rate of star 
formation in galaxies does not need to be closely correlated with the 
rate of cluster formation.  In fact, Hodge (1998) finds that the rate 
of cluster formation, normalized to similar rates of star formation, is 
presently
$\geq 600$ times greater in the LMC than it is in IC  1613.  The observation 
(Whitmore \& Schweizer 1995) that star cluster are being formed very 
actively in the violently interacting galaxies NGC 4038/39 (``the 
antennae'') suggests that strong shocks might favor the formation of 
clusters.  Finally, it is of interest to note that the rate of star and 
cluster formation in the SMC (Mighell, Sarajedini \& French 1999) 
appears to have proceeded at a more-or-less constant rate over the last 
$\sim 10$ Gyr.  This shows that tidal interactions between the Clouds 
cannot be invoked to account for the LMC starbursts that started $\sim 4$ 
Gyr ago.

\subsection{Early Star Formation}

Olsen (U. Washington) et al.~used deep HST observations to show that 
the LMC globular clusters NGC 1835, NGC1898, NGC 1916, NGC 2005, and 
NGC 2019 have ages that differ by $\leq 1$ Gyr.  Similar results were 
obtained for NGC 1466, NGC 2259 and Hodge 11 by Johnson (UCSC) et al.  
Furthermore the absolute ages of these clusters are found to be similar 
to those of the Galactic globulars M~3, M~5 and M~55.  These results 
suggest that the Population II component of the Large Cloud formed 
during a short, but intense, burst of star and cluster formation.  On 
the other hand deep HST observations of clusters in the SMC by Mighell 
(NOAO) et al.~show a rather more gradual onset of cluster formation, 
with the oldest SMC cluster NGC 121 having an age of only $10.6 \pm 0.7$ 
Gyr.  Hesser (DAO) et al.~also used HST observations to show that the 
luminous outer halo Galactic globular NGC 2419, which is located at 
R$_{\mathrm Gc} \sim 100$ kpc, has an age similar to that of the globular 
clusters in the LMC and in the main body of the Galactic halo.  
On the other hand they 
find that the fainter outer Galactic halo clusters Palomar 3, Palomar 4 
and Eridanus have lower ages that are similar to that of NGC 121 in the 
SMC.  The picture that emerges from these results is one in which 
massive globular clusters formed almost simultaneously in the LMC, and 
the inner and outer halo of the Galaxy, i.e. throughout a region with a 
radius of $\sim 100$ kpc.  Less luminous massive clusters formed in the 
outer Galactic halo and in the SMC during the next few Gyr.

Grebel (Lick Obs) et al.~strengthened observational data which show 
that both Magellanic Clouds shrank as they evolved, with the oldest 
Population II component occupying a larger volume than younger stars 
belonging to Population I.  Van den Bergh (2000) has noted a similar 
effect in all other Local Group dwarf irregulars in which detailed 
population studies have so far been made.

\section{The Distances to the Magellanic Clouds}

\subsection{Distance to the Large Cloud}

Distance determinations to the LMC prior to 1996 have been reviewed by 
Westerlund  (1997).  An unweighted mean of these data yields a mean 
distance modulus  $<(m-M)_{\mathrm o}> = 18.48 \pm 0.04$.  
A tabulation of 16 more 
recent distance determinations, which was handed out at the Symposium 
by van den Bergh, yields an unweighted mean value 
$<(m-M)_{\mathrm o}> = 18.50 \pm 0.04$.  
Three of the determinations listed in this table give results 
that are inconsistent with this mean:  Using {\em Hipparcos} parallaxes Feast 
\& Catchpole (1997) find 
$(m-M)_{\mathrm o} = 18.70 \pm 0.10$.  However, after 
applying Lutz-Kelker corrections Oudmaijer et al.~(1998) obtain a 
smaller value 
$(m-M)_{\mathrm o} = 18.56 \pm 0.08$.  The appropriateness of these 
corrections was, however, disputed by Feast.  A second discordant 
distance to the LMC was found by Udalski et al.~(1998), who obtained 
$(m-M)_{\mathrm o} = 18.08 \pm 0.03 \pm 0.12$ (systematic) from the mean magnitude of 
red clump  giants in the LMC.  A rediscussion of these results (Cole 
1998) yields a larger modulus $(m-M)_{\mathrm o} = 18.36 \pm 0.17$.  Even more 
recently Paczy\'{n}ski (1998) has questioned the validity of using the 
magnitudes of clump stars as distance indicators.  Finally Reid (1997) 
has obtained a discordant modulus of 
$(m-M)_{\mathrm o} = 18.71 \pm 0.06$ by fitting 
the RR Lyrae variables in the LMC globulars NGC  1466 and NGC 2257 to 
those in the Galactic globular cluster NGC 6397.  From this fit Reid 
obtains $(m-M)_{\mathrm o} = 18.71 \pm 0.06$.  The weak link in this chain might be 
the fit of the main sequence of NGC 6397 to the {\em Hipparcos} parallaxes of 
nearby subdwarfs.

Perhaps the most direct determinations of the LMC distance are those 
which are purely geometrical in nature.  At the meeting Pritchard (Mt. 
John Obs.) et al.~presented observations of the detached early-type 
eclipsing binary HV 2274 from which they obtained 
$(m-M)_{\mathrm o} = 18.44 \pm 0.07$.  Five additional detached 
early-type binaries in the LMC have yet 
to be observed and might strengthen this distance determination.  A 
second purely geometrical method of distance determination is based on 
ultraviolet observations of the SN 1987A ring.  Panagea (STScI/ESA) 
showed that these data yield $(m-M)_{\mathrm o} = 18.55 \pm 0.05$ for the supernova, 
and $(m-M)_{\mathrm o} = 18.58 \pm 0.05$ for the centroid of the  LMC.  In view of 
these results it seems safest to continue to use the canonical value 
$(m-M)_{\mathrm o} = 18.5 \pm 0.1$, corresponding to D(LMC) $= 50$ kpc.

\subsection{Distance to the Small Cloud}

The SMC distance determinations prior to 1996 have been reviewed by 
Westerlund (1997).  An unweighted mean of these data yields 
$<(m-M)_{\mathrm o}> = 19.94 \pm 0.05$.  
After excluding the red clump modulus of Udalski et 
al.~(1998), for the reasons discussed above, one finds a mean 
unweighted value of $<(m-M)_{\mathrm o}> = 18.85 \pm 0.04$ for the SMC from distance 
determinations that have been  published  recently.  It is, however, a 
source of grave concern that the four Cepheid-based determinations 
yield $(m-M)_{\mathrm o} = 18.93 \pm 0.03$, which is inconsistent with 
$(m-M)_{\mathrm o} = 19.73 \pm 0.02$ that is derived from five distance 
determinations based 
on RR Lyrae variables.  Kunkel (private communication) has suggested 
that this difference may be due to the great 
depth along the line of sight of the SMC.  
Perhaps many of the young Cepheids in the Small Cloud are located in 
the tidal tail behind main body of the SMC.  In summary it appears that 
the distance modulus of the Small Cloud is probably 
$(m-M)_{\mathrm o} = 18.85 \pm 0.1$ (corresponding to D $= 59$ kpc), with the 
caveat that  young stars might, on average, be more distant than the 
main body of this galaxy.  
Caldwell \& Coulson have pointed out that the distance modulus of the 
Wing of the SMC is probably $\sim 0.3$ mag smaller than that of the main 
body of the Small Cloud.

\section{The Orbit of the Magellanic Clouds}

\subsection{Evidence for Recent Interactions}

The discovery of the Magellanic Stream by Mathewson, Cleary \& Murray 
(1974) provided the first dramatic evidence for a strong tidal 
interaction between the Magellanic Clouds $\sim 1.5$ Gyr ago (Gardiner \& 
Noguchi 1996).  A second encounter 0.2-0.3 Gyr ago is believed by 
Demers (U. Montreal) \& Kunkel (OCIW) to have resulted in the formation 
of the Bridge between the LMC and SMC, and of the tidal tail behind the 
Small Cloud.  Venn (Macalester College) found [Fe/H] $\sim -1.0$ for 
supergiants in the Bridge.  This low metallicity suggests that gas in 
the Bridge was mainly drawn from the SMC.  At the Symposium Putman 
(MSSSO) et al.~reported the discovery of a narrow continuous gaseous 
tail, which leads the direction of motion of the Clouds, i.e. in the 
direction opposite to that of the Stream.  Majewski (U. Virginia) et 
al.~have searched for stars associated with the Magellanic Stream and 
reported that they have found a number of giant stars concentrated at 
distances expected for tidal debris from the Magellanic Clouds.  Wakker 
(U. Wisconsin) et al.~suspect that the high velocity cloud 287+22+240 
may represent metal-poor material that originated in the Magellanic 
Stream.

\subsection{The Orbit(s) of the Magellanic Clouds}

Byrd et al.~(1994) have modelled the interactions between the members 
of the Local Group from which they concluded that the Magellanic Clouds 
may have left the neighborhood of the Andromeda galaxy $\sim 10$ Gyr ago, 
and were subsequently captured by the Galaxy $\sim 6$ Gyr ago.  Simulations 
by Sawa (Aichi U.) et al.~also suggest that the LMC and SMC have formed 
a bound pair for $\sim 15$ Gyr.  On the other hand detailed simulations by 
Li (U. Wyoming) \& Thronson (NASA) show that the Small Cloud lost so 
much matter during its two most recent interactions with the Large 
Cloud that it cannot have survived many such interactions.  They 
therefore concluded that the LMC and SMC must have captured each other 
fairly recently.  Only greatly improved proper motions for the Clouds 
will place significant constraints on the evolution of their orbital 
history.  Unfortunately the presently available data on their proper 
motions, which are listed in Table 1, show inconsistent motions.

\begin{table}
\caption{Proper Motion of LMC in Milliarcseconds}
\begin{center}
\begin{tabular}{ccll}
\tableline
\tableline
$\mu_{\alpha}\cos{\delta}$ & $\mu_{\delta}$ &    Method      & Reference \\
\tableline
$+1.20 \pm 0.28$ & $+0.26 \pm 0.27$ & Relative to galaxies & Jones et al.~(1994)\\
$+1.94 \pm 0.29$ & $-0.14 \pm 0.36$ & Hipparcos            & Kroupa \& Bastian (1997) \\
$+1.6 \pm 0.2$    & $+3.0 \pm 0.2$   & Relative to quasars  & Anguita\tablenotemark{1} (1998)\\
\tableline
\tablenotetext{1}{These proceedings}
\end{tabular}
\end{center}
\end{table}

\section{Gravitational Lensing}

Paczy\'{n}ski (1986) wrote that ``Monitoring the brightness of a few 
million stars in the Magellanic Clouds over a time scale between 2 hr 
and 2 yr may lead to the discovery of ``dark halo'' objects in the mass 
range $10^{-6} - 10^{+2}$ M$_{\odot}$ or it may put strong upper limits 
on the number of 
such objects.''  This prediction has been brilliantly confirmed by 
observations of the EROS, MACHO and OGLE consortia.  Particularly 
exciting results, on the lensing event that took place in the SMC less 
than a month before the symposium, were reported by Alves (LLNL) et al. 
 The lightcurve of this event showed that it was produced by a binary.  
Such data on microlensing of binaries can break the degeneracy between 
mass, location, and transverse velocity, that occurs in the standard 
gravitational microlensing model for single stars.  In the case of the 
June 1998 event seen in the direction of the SMC, it was found that the 
observed proper motion of the lensing object is so small that there is 
only a 0.15\% probability that it was produced by a foreground object in 
the Galactic halo.  Previous observations of lensing events in the 
direction of the LMC had already excluded objects in the mass range 
$10^{-7}$ M$_{\odot}$ to 1 M$_{\odot}$ as significant contributors to 
the mass of the Galactic dark halo.

The EROS, MACHO and OGLE surveys have also produced a massive and 
homogeneous database on variable stars in the Clouds of Magellan.  
Investigations based on these new data on variable stars were reported 
by Welch (McMaster), Alves (LLNL) et al., Marquette (Inst. d'Ap.) and 
Bono (Trieste) et al.

\section{Conclusions}

\begin{itemize}
\item{Geometrical distance determinations, based on observations of SN 
1987A and of the detached eclipsing binary HV 2274, yield distance 
moduli of $(m-M)_{\mathrm o} = 18.58 \pm 0.05$ and 
$(m-M)_{\mathrm o} = 18.44 \pm 0.07$, 
respectively for the Large Magellanic Cloud.  These values are both 
compatible with the canonical value $(m-M)_{\mathrm o} = 18.5 \pm 0.1$, which 
corresponds to a distance of 50 kpc.}

\item{The great burst of cluster formation that started in the LMC 3-5 Gyr 
ago is only weakly reflected in the rate at which field stars were 
formed.  This strongly suggests that the rate of cluster formation is 
not a good diagnostic for the overall rate of star formation.  The 
observation that the present rate of cluster formation, normalized to 
the rate of star formation in the LMC, is more than two orders of 
magnitude greater than it is in the Local Group dwarf irregular IC 1613 
supports this conclusion.}

\item{Tidal interactions between the LMC and SMC that occurred $\sim 0.2$ Gyr 
and $\sim 1.5$ Gyr ago produced the Bridge and the Magellanic Stream, 
respectively.  It is presently not clear if the LMC and SMC were closely 
bound between 3 Gyr and 13 Gyr ago.  Improved proper motions are 
urgently required to constrain their orbital history.}

\item{Observations of microlensing events strongly suggest that they are 
not produced by objects located in the Galactic halo.  The enormous 
data base provided by the EROS, MACHO and OGLE consortia is proving to 
be a gold mine for the study of variable stars in the Clouds of 
Magellan.}

\end{itemize}

\end{document}